\documentclass[a4paper]{article}

\usepackage{siunitx}
\usepackage{xcolor}

\newcommand{\norm}[1]{\left\lVert#1\right\rVert}

\usepackage{float}
\usepackage{aliascnt}
\newaliascnt{eqfloat}{equation}
\newfloat{eqfloat}{h}{eqflts}
\floatname{eqfloat}{Equation}

\newcommand*{\ORGeqfloat}{}
\let\ORGeqfloat\eqfloat
\def\eqfloat{%
  \let\ORIGINALcaption\caption
  \def\caption{%
    \addtocounter{equation}{-1}%
    \ORIGINALcaption
  }%
  \ORGeqfloat
}

\newtheorem{theorem}{Theorem}
\usepackage{authblk}
\usepackage{multirow}
\usepackage{amsmath} 
\usepackage[skip=0.5\baselineskip]{caption}
\usepackage{url}

\usepackage{mathtools}
\usepackage{pdflscape}

\usepackage{array}
\usepackage{newtxtext,newtxmath}

\usepackage{breqn}

\begin{document}
\title{Hypothesis testing for matched pairs with missing data by maximum mean discrepancy: An application to continuous glucose monitoring}

\author{MARCOS MATABUENA$^{\ast a}$, PAULO F\'{E}LIX$^{a}$, MARC DITZHAUS$^b$, JUAN  VIDAL$^{a,c}$,\\ FRANCISCO GUDE$^{d}$\\[4pt]
\textit{$^a$Centro Singular de Investigaci\'on en Tecnolox\'ias Intelixentes (CiTIUS), Universidade de Santiago de Compostela, Spain}\\
\textit{$^b$Otto-von-Guericke-Universit\"at Magdeburg, Fakult\"at f\"ur Mathematik (FMA), Institut f\"ur Mathematische
Stochastik (IMST)}\\
\textit{$^c$Departamento de Electr\'{o}nica e Computaci\'{o}n, Universidade de Santiago de Compostela, Spain}\\
\textit{$^d$Unidad de Epidemiolog\'{i}a Cl\'{i}nica, Hospital Cl\'{i}nico Universitario de Santiago, Spain}\\[2pt]
{marcos.matabuena@usc.es}}

\markboth%
{M. Matabuena and others}
{Hypothesis testing in the presence of complex paired missing data}

\maketitle

\footnotetext{To whom correspondence should be addressed.}

\begin{abstract}{
A frequent problem in statistical science is how to properly handle missing data in matched paired observations. There is a large body of literature coping with the univariate case. Yet, the ongoing technological progress in measuring biological systems raises the need for addressing more complex data, e.g., graphs, strings and probability distributions, among others. In order to fill this gap, this paper proposes new estimators of the maximum mean discrepancy (MMD) to handle complex matched pairs with missing data. These estimators can detect differences in data distributions under different missingness mechanisms. The validity of this approach is proven and further studied in an extensive simulation study, and results of statistical consistency are provided. Data from continuous glucose monitoring in a longitudinal population-based diabetes study are used to illustrate the application of this approach. By employing the new distributional representations together with cluster analysis, new clinical criteria on how glucose changes vary at the distributional level over five years can be explored.}

\text{Paired missing data; Kernel methods; Distributional representations; Diabetes Mellitus} \\
\end{abstract}

\section{Introduction}
\label{sec1}

A common experimental design in clinical studies, especially longitudinal ones, is the matched pairs design where observations are made from the same subjects under two different conditions, often at two points in time, before and after treatment, or after a fixed period of time from the baseline. Revealing the possible dependence structure of the observations is a major goal. However, a typical issue when dealing with paired data is the occurrence of missing data. The challenge is to exploit the available data to perform valid inference.

The literature on matched pairs with missing data has primarily focused on one-dimensional, continuous, discrete or ordinal variables, aimed at detecting changes in location/mean \cite{ekbohm1976comparing,guo2017comparative,martinez2013hypothesis,xu2012accurate}, scale/variance \cite{derrick2018tests}, and distribution \cite{gaigall2020testing}. Some of the proposals apply multiple imputation techniques \cite{akritas2002nonparametric,akritas2006nonparametric,verbeke2009}, but they often require large sample sizes for being correct. Other proposals rely on specific model assumptions such as symmetry or bivariate normality \cite{ekbohm1976comparing,samawi2011,xu2012accurate}, but they exhibit a non-robust behavior against deviations. The common approach recently adopted in literature results from combining in a non parametric approach separate test statistics for the paired and unpaired observations, by using either weighted test statistics \cite{amro2017permuting,fong2018rank,gaigall2020testing,konietschke2012ranking,martinez2013hypothesis,samawi2014notes}, a multiplication combination test \cite{amro2019}, or combined p-values \cite{amro2021asymptotic,kuan2013,yu2012permutation,qi2019}

The recent scientific and technological progress in measuring biological processes has enabled monitoring of patient's condition with a growing level of detail and complexity. Thus, beyond the ongoing identification of univariate biomarkers, new complex data structures are being incorporated into the analysis, as is the case of population ages and mortality distributions \cite{bigot2017}, distributions of functional connectivity patterns in the brain \cite{dubey2020functional,petersen2016functional}, post-intracerebral hemorrhage hematoma densities \cite{petersen2021}, graph-based representations of connectivity and functional brain activity \cite{takerkart2014}, and glucose distributions from continuous monitoring \cite{matabuena2021glucodensities}. 

The aim of the present paper is to provide a statistical test for matched pairs with missing data which does not require any parametric assumptions and uses all observations available. We propose new maximum mean discrepancy (MMD) estimators to achieve this aim \cite{gretton2012kernel}. The energy distance and the MMD are two equivalent statistical metrics with the ability to detect distributional differences between random samples \cite{szekely2013energy,shen2021exact}. Moreover, MMD-based statistics can also be seen as a natural generalization of the ANOVA test to cases where the distributions are not necessary Gaussian \cite{rizzo2010disco}. MMD overcomes Gaussian assumptions by representing distances between distributions as distances between mean embeddings in a reproducing kernel Hilbert space (RKHS). MMD has been successfully applied to independence testing \cite{szekely2007measuring}, two-sample testing \cite{gretton2012kernel}, survival analysis \cite{fernandez2019maximum}, or clustering analysis \cite{francca2021kernel}.

Besides conducting an extensive simulation study, the new testing procedures are applied to the AEGIS diabetes dataset, resulting from a longitudinal population-based study \cite{gude2017}. This dataset includes data from continuous glucose monitoring (CGM), performed at the beginning of the study and five years later. Importantly, there is a substantial loss to follow up. A distributional representation of glucose concentration summarizes several days of monitoring, providing a personal signature of glucose homeostasis \cite{matabuena2021glucodensities}. The present approach allows us to address some interesting questions related to the possible changes in CGM profile with ageing, or the relation between obesity and diabetes. Furthermore, an adaption of a previous clustering method to matched pairs with missing data allows us to find out specific patient phenotypes, with potential applications in patient stratification \cite{francca2021kernel}.

The rest of this paper is outlined as follows. In Section \ref{sec:glucodensities} we provide a motivation for the new methods from the distributional representation of CGM data. In Section \ref{sec:methods} we define the problem in general terms and introduce the statistical model based on the MMD metric, providing weighted test statistics for dealing with missing data under MCAR mechanism (Section \ref{sec:mcar}) and under MAR mechanism (Section \ref{sec:mar}). A proof presenting theoretical guarantees of the proposed methods is delivered in the appendix. In Section \ref{sec:kernel_choice} the choice of kernel functions and corresponding hyperparameters is discussed. Then we present the results of an extensive simulation study in Section \ref{sec:simulation_study}. In Section \ref{sec:cluster} a previous clustering method is adapted to missing data under the MAR mechanism. We present in Section \ref{sec:aegis} some applications of both hypothesis testing and clustering analysis to the AEGIS study, by exploiting the distributional representation of CGM data. We close with a discussion in Section \ref{sec:discussion}.

\section{Motivation: a distributional representation of continuous glucose monitoring data}
\label{sec:glucodensities}

Distributional data analysis is a novel methodology that has proved successful to manage biosensor data in different settings such as the connectivity analysis of the brain network \cite{petersen2016functional}, the diabetes management \cite{matabuena2021glucodensities}, and the physical activity analysis \cite{ghosal2021distributional,matabuena2021distributional}.

In a previous paper \cite{matabuena2021glucodensities}, we introduce a novel distributional representation for CGM data, termed glucodensity, which allows us to obtain a functional profile of patient glucose homeostasis. Glucodensity is a natural extension of Time in Range (TIR) metrics, that measures the proportion of time a person spends with their blood glucose levels within the target range of $70-180$ mg/dL  \cite{battelino2019clinical,beck2019validation}. Although very intuitive, TIR metrics have two main disadvantages: first, the range fits poorly depending on the characteristics of the population examined; second, there is a loss of information caused by the discretization of the recorded data into intervals. Instead, glucodensity effectively measures the proportion of time each individual spends at a specific glucose concentration. Previous results for glucodensities show a better predictive performance as compared to common diabetes biomarkers.

\begin{figure}[h]
	\centering
	\includegraphics[width=0.8\linewidth]{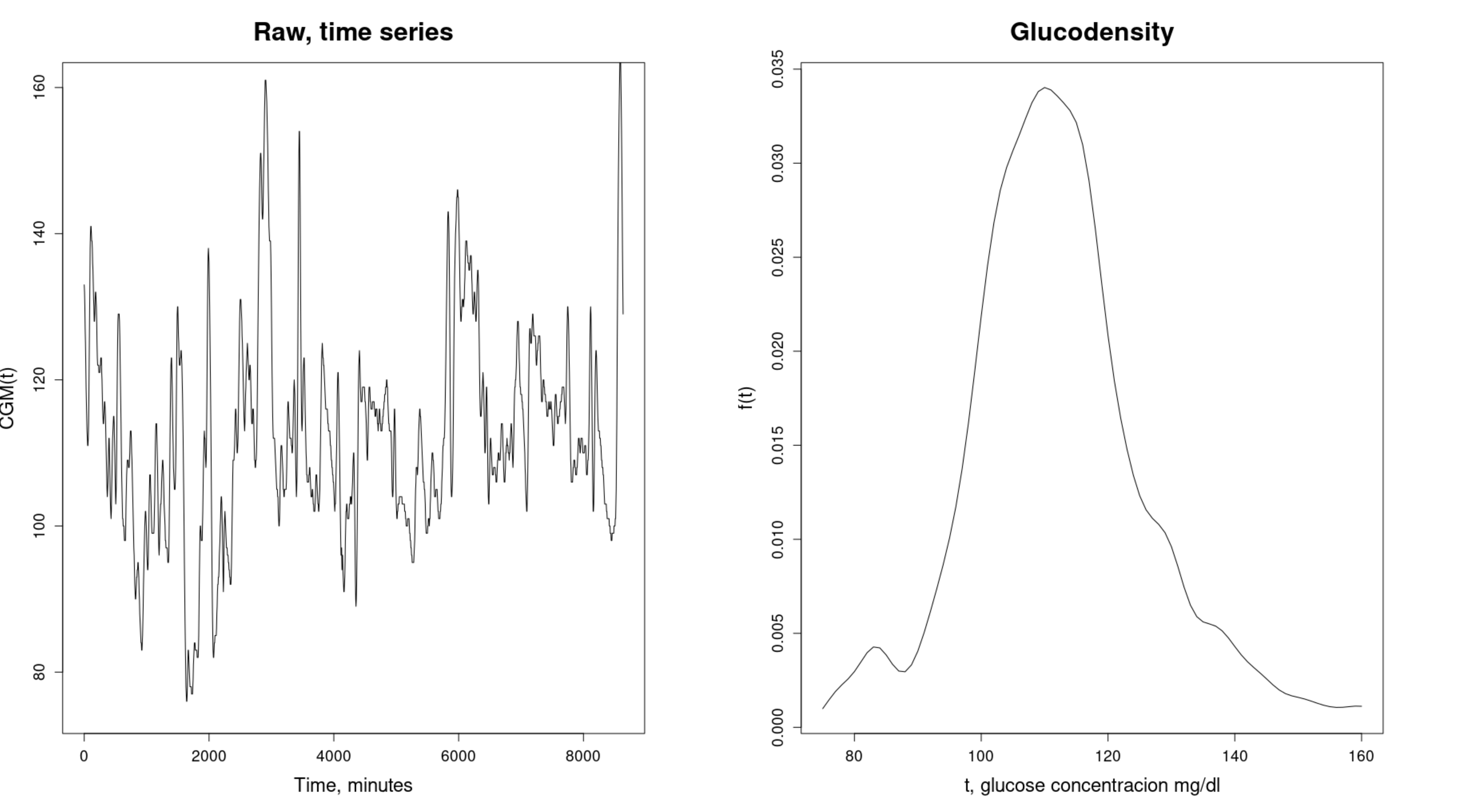}
	\caption{Left: The CGM recording from a normoglycemic patient. Right: The corresponding glucodensity.
	}
	\label{fig:gluco_overview}
\end{figure}

Given a series of CGM data $\{Y_j\}_{j=1}^m$, the glucodensity can be modeled as a probability density function $f(\cdot)$ that can be approached by kernel density estimation:
\begin{equation}
 \hat f(y)=\frac{1}{m}\sum _{j=1}^{m} \frac{1}{h} K\left(\frac{Y_j-y}{h}\right),
\end{equation}
where $h>0$ is the smoothing parameter and $k(\cdot)$ denotes a non-negative real-valued integrable function (Figure \ref{fig:gluco_overview}). 

Let $\mathcal{D}$ be the space of probability density functions $f$ such that $\int _{\mathbb{R}}u^2f(u)du <\infty$. In order to measure the difference between two glucodensities, $f$ and $g$, a metric on $\mathcal{D}$ is required. We use the $2-$Wasserstein distance:
\begin{equation}
d^2_{\mathcal{W}_2}(f,g)= \int_{0}^{1} \left|Q_{f}\left(t\right)-Q_{g}\left(t\right)\right|^{2}dt,  \quad f, g\in \mathcal{D},
\label{eq:wasserstein}
\end{equation}
where $Q_{f}$ and $Q_{g}$ denote the corresponding quantile functions. The $2$-Wasserstein distance in (\ref{eq:wasserstein}) depends only on quantile functions, so we can approximate it by computing empirical quantile functions $\hat{Q}_f=\hat{F}^{-1}$ and $\hat{Q}_g=\hat{G}^{-1}$, from the corresponding empirical cumulative distribution functions $\hat{F}$ and $\hat{G}$.

\begin{figure}[ht!]
	\centering
	\includegraphics[width=0.6\linewidth]{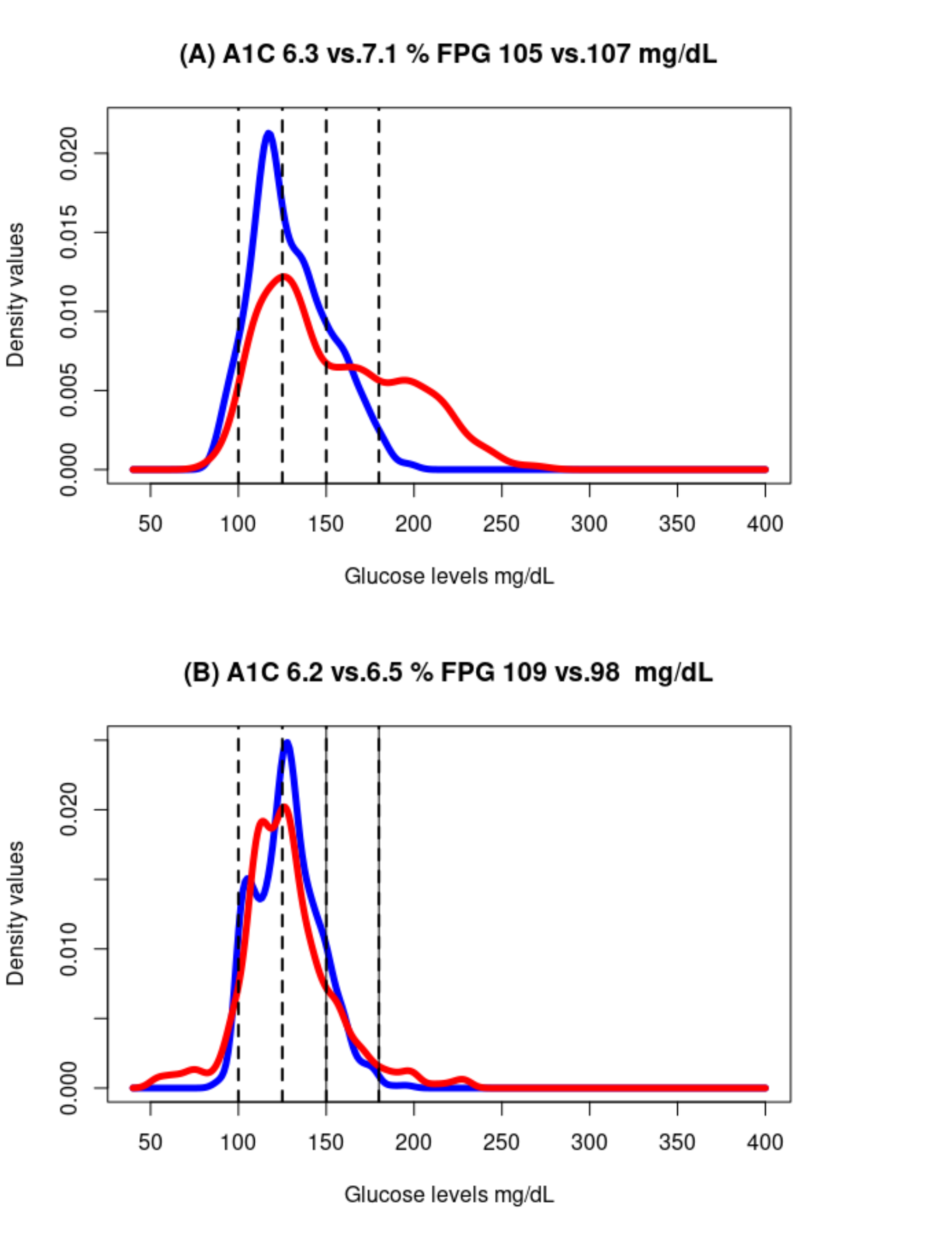}
	\caption{Glucodensity changes in prediabetic patients (blue) who develope diabetes after $5$ years (red).}
	\label{fig:glucodensity_changes}
\end{figure}

Figure \ref{fig:glucodensity_changes} contains an example of the glucodensity representation for the continuous glucose monitoring performed on three different individuals, both in a prediabetes and later diabetes status. This figure immediately poses the challenge of defining new statistical methods to compare two sets of glucodensity measurements to assess whether some population statistics differ. This can be useful to compare the glucose homeostasis before and after a treatment, or after a certain period of time.

\section{Hypotheses and statistics}
\label{sec:methods}

Let $\mathcal{D}$ be a separable Hilbert space and $\left(X_{1},X_{2}\right)^\top \in \mathcal{D}^2$ a random pair representing two different measurements on a subject at two different time points. Let us consider a general matched pairs design given by i.i.d. random variables
\begin{equation}
 \mathbf{X}_j=\begin{pmatrix} X_{1j} \\ X_{2j} \end{pmatrix}, ~ j=1,...,n.
\end{equation}
To continue the example of our motivation, $X_{1j}$ can represent the glucodensity at the beginning of a certain study for the j-th patient, and $X_{2j}$ the glucodensity at the end of the study for the same patient. Let assume that both $\{X_{1j}\}_{j=1}^n$ and $\{X_{2j}\}_{j=1}^n$ are drawn from probability measures $P_1$ and $P_2$, respectively. We are interested in testing the equality of distributions as null hypothesis $H_{0}:\{P_{1} = P_{2}\}$ against the alternative $H_{1}:\{ P_{1} \neq P_{2}\}$, i.e., to check whether there are systematical differences between the outcomes at different time points. 

\subsection{Missing Completely at Random (MCAR) mechanism}
\label{sec:mcar}

When some of the elements of the matched pairs are missing completely at random the available data can be sorted as:
\begin{equation}
\mathbf{X}=\underbrace{\begin{pmatrix}   X_{11} \\ X_{21} \end{pmatrix} \cdots    \begin{pmatrix}   X_{1n_1} \\ X_{2n_1} \end{pmatrix}}_{
\substack{\text{Complete data $\mathbf{X}^{\text{com}}$} \\ \text{$n_1$ observations}}}
\hspace{0.2cm}  \underbrace{\begin{pmatrix}   X_{1n_{1+1}} \\  -  \end{pmatrix} \cdots   \begin{pmatrix}   X_{1n_1+n_2} \\ -  \end{pmatrix}}_{
\substack{\text{Incomplete data $\mathbf{X}^{\text{inc}}_1$} \\ \text{$n_2$ observations}}} \hspace{0.2cm}  \underbrace{\begin{pmatrix} -  \\  X_{2 n_1+n_2+1}   \end{pmatrix} \cdots   \begin{pmatrix}   -  \\ X_{2 n_1+n_2+n_3}  \end{pmatrix}}_{
\substack{\text{Incomplete data $\mathbf{X}^{\text{inc}}_2$} \\ \text{$n_3$ observations}}}, \\
\end{equation}
where $n=n_1+n_2+n_3$. For ease of notation, we denote $\mathbf{X}^{\text{com}}_1=\{X_{1j}\}_{j=1}^{n_1}$, $\mathbf{X}^{\text{com}}_2=\{X_{2j}\}_{j=1}^{n_1}$, $\mathbf{X}^{\text{inc}}_1=\{X_{1j}\}_{j=n_1+1}^{n_1+n_2}$ and $\mathbf{X}^{\text{inc}}_2=\{X_{2j}\}_{j=n_1+n_2+1}^{n}$. Additionally, a missingness status variable can be defined  
 $\delta_{ij}\in \{0,1\}$, $i=1,2$, $j=1,\dots, n$, so $\delta_{ij}=1$ if the element is missing and $\delta_{ij}=0$ otherwise. 

A natural way of testing the equality of distributions is measuring the distance between them. We propose two test statistics: $\mathcal{T}_1$ for the complete data sets $\mathbf{X}^{\text{com}}_1$ and $\mathbf{X}^{\text{com}}_2$, and $\mathcal{T}_2$ for the incomplete data sets $\mathbf{X}^{\text{inc}}_1$ and $\mathbf{X}^{\text{inc}}_2$, which are then combined in one weighted test statistic:
\begin{equation}
\mathcal{T}(\mathbf{X})= \alpha \mathcal{T}_1(\mathbf{X}^{\text{com}}_1,\mathbf{X}^{\text{com}}_2)+(1-\alpha) \mathcal{T}_2(\mathbf{X}^{\text{inc}}_1,\mathbf{X}^{\text{inc}}_2),
\label{eq:test_statistic}
\end{equation}
for some weighting parameter $\alpha \in \left[0,1\right]$. Both $\mathcal{T}_1$ and $\mathcal{T}_2$ are based on the maximum mean discrepancy (MMD) to measure the empirical distance between the marginal distributions \cite{gretton2012kernel}. Let $k: \mathcal{D}\times \mathcal{D} \to \mathbb{R}^{+}$ be a symmetric definite positive kernel. The existence of a dot product space $\mathcal{H}$ and feature mapping $\phi: \mathcal{D} \to \mathcal{H}$ is guaranteed, such that $k(X,X')=\langle \phi(X),\phi(X')\rangle_{\mathcal{H}}$. A reproducing kernel of $\mathcal{H}$ is a kernel function that satisfies (1) $\forall X\in \mathcal{D}$, $k(\cdot,X)\in \mathcal{H}$, and (2) $\forall X\in \mathcal{D}$, $\forall g\in \mathcal{H}$, $\langle g,k(\cdot,X) \rangle_{\mathcal{H}}=g(X)$. $\mathcal{H}$ is then said to be a reproducing kernel Hilbert space (RKHS). Kernel mean embedding results from extending the mapping $\phi$ to the space of probability distributions by representing each distribution as a mean function $\phi(F) = \mathbf{E}[k(\cdot,X)] = \int_{\mathcal{D}} k(\cdot,X)dP$. The kernel mean embedding can be empirically estimated by $\widetilde{\phi}= \frac{1}{n}\sum_{i=1}^{n}k\left(\cdot, X\right)$. Then, we can measure the distance between random samples as follows: 
\begin{align*}
	   \mathcal{T}_1\left(\mathbf{X}^{\text{com}}_1,\mathbf{X}^{\text{com}}_2\right)&=  \norm{\widetilde{\phi}^{\text{com}}_{1}-\widetilde{\phi}^{\text{com}}_{2}}_{\mathcal{H}}^{2}\\
	   &=  \langle \frac{1}{n_1}\sum_{i=1}^{n_1}k\left(\cdot, X_{1i}\right)-\frac{1}{n_1}\sum_{i=1}^{n_1}k\left(\cdot, X_{2i}\right), \frac{1}{n_1}\sum_{i=1}^{n_1}k\left(\cdot, X_{1i}\right)-\frac{1}{n_1}\sum_{i=1}^{n_1}k\left(\cdot, X_{2i}\right) \rangle  \\
	   &=\frac{1}{n_1^{2}}  \sum_{i=1}^{n_1}\sum_{j=1}^{n_1} k\left(X_{1i},X_{1j}\right)+\frac{1}{n_1^{2}}  \sum_{i=1}^{n_1}\sum_{j=1}^{n_1} k\left(X_{2i},X_{2j}\right)-\frac{2}{n_1^{2}} \sum_{i=1}^{n_1}\sum_{j=1}^{n_1}k\left(X_{1i},X_{2j}\right).
\end{align*}
Analogously,
\begin{align*}
	\mathcal{T}_2\left(\mathbf{X}^{\text{inc}}_1,\mathbf{X}^{\text{inc}}_2\right) &= \norm{\widetilde{\phi}^{\text{inc}}_{1}-\widetilde{\phi}^{\text{inc}}_{2}}_{\mathcal{H}}^{2} \\ 
	&=\frac{1}{n_2^{2}} \sum_{i=n_1+1}^{n_1+n_2} \sum_{j=n_1+1}^{n_1+n_2} k\left(X_{1i}, X_{1j}\right)+\frac{1}{n_3^{2}} \sum_{i=n_1+n_2+1}^{n_1+n_2+n_3} \sum_{j=n_1+n_2+1}^{n_1+n_2+n_3} k\left(X_{2i}, X_{2j}\right)\\
	&\phantom{=}\,\,-\frac{2}{n_2n_3}  \sum_{i=n_1+1}^{n_1+n_2} \sum_{j=n_1+n_2+1}^{n_1+n_2+n_3} k\left(X_{1i},X_{2j}\right).
\end{align*}
Importantly, for the class of {\em characteristic} kernels, the embeddings are injective, and hence $\norm{P_{1}-P_{2}}_{\mathcal{H}}^{2} = 0$, if and only if $P_1 = P_2$ \cite{sriperumbudur2011universality}. 

In order to calibrate the tests under the null hypothesis it should be pointed out that both $\mathcal{T}_1$ and $\mathcal{T}_2$ do not follow a free asymptotic distribution. The empirical estimate of MMD is a one-sample V-statistics and hence asymptotic distribution is difficult to obtain due to the degeneracy of V-statistics, which incorporates a correlation structure for the complete paired observations $\mathbf{X}^{\text{com}}$ \cite{gretton2012kernel}. To address this issue we propose a wild bootstrap procedure for the first $n_1$ observations, while the remaining $n_2+n_3$ observations can be properly handled by permutations methods, that can achieve an exact type I error control. For each $b=1,\dots,B$, it proceeds as follows:

\begin{enumerate}
\item For the first $n_1$ complete paired observations, take random weights $w^b_{i}$, $i=1,\dots, n_1$, with  
\begin{align*}
w^b_{i}= e^{-1/l_{n_1}}w^{b}_{i-1}+\sqrt{1-e^{-2/l_{n_1}}}\epsilon_i, 
\end{align*}
where $w^b_0,\epsilon_1,\cdots, \epsilon_{n_1}$ are independent standard normal variables, and $l_{n_1}$ is a bootstrap parameter used to mimic the dependence structure, such that $l_{n_1}= o\left(n_1\right)$ but $ \lim_{n_1\to \infty} l_{n_1}=\infty$. Then,
\begin{align*}
\mathcal{T}^{b}_1\left(\mathbf{X}^{\text{com}}_1,\mathbf{X}^{\text{com}}_2\right)=   \frac{1}{n_1^{2}}  \sum_{i=1}^{n_1}\sum_{j=1}^{n_1}  w^b_{i} w^b_{j} \left[ k\left(X_{1i},X_{1j}\right)+ k\left(X_{2i},X_{2j}\right)-2k\left(X_{1i},X_{2j}\right) \right].
\end{align*}
\item The remaining $n_2+n_3$ observations belonging to $\mathbf{X}^{\text{inc}}_1$ and $\mathbf{X}^{\text{inc}}_2$ are randomly permuted, i.e. each observation is randomly assigned to new $\mathbf{X}^{\text{inc},\pi}_1$ or $\mathbf{X}^{\text{inc},\pi}_2$ sets, resulting in new $\mathcal{T}_2^b(\mathbf{X}^{\text{inc},\pi}_1,\mathbf{X}^{\text{inc},\pi}_2)$.
\item Then, calculate   
\begin{align*}
    \mathcal{T}^{b}=   \alpha \mathcal{T}^{b}_1 \left(\mathbf{X}^{\text{com}}_1,\mathbf{X}^{\text{com}}_2 \right)+(1-\alpha)\mathcal{T}_2^b\left(\mathbf{X}^{\text{inc},\pi}_1,\mathbf{X}^{\text{inc},\pi}_2\right)
\end{align*}
\end{enumerate}

Finally, return $p$-value$= \frac{1}{B} \sum_{b=1}^{B} 1\{ \mathcal{T}^{b}  \geq \mathcal{T}(\mathbf{X})\}$.

\begin{theorem}
\label{pre:1}
Let $\mathbf{X}^{\text{com}}=\{\left(X_{1i}, X_{2i}\right)^{\top}\}^{n_1}_{i=1}$ be a set of i.i.d. complete paired samples, and $\mathbf{X}^{\text{inc}}_1=\{X_{1i}\}^{n_1+n_2}_{i=n_1+1}$, and $\mathbf{X}^{\text{inc}}_2=\{X_{2i}\}^{n_1+n_2+n_3}_{i=n_1+n_2+1}$ two sets of i.i.d. incomplete paired samples. Let suppose that $n_1/\left(n_1+n_2+n_3\right)\to\kappa_1\in \left(0,1\right)$ and $n_2/(n_2+n_3)\to \kappa_2 \in \left(0,1\right)$ as $n_1,n_2,n_3\to \infty$; then, the test statistic given by (\ref{eq:test_statistic}) is consistent against the alternative $H_{1}: \{P_{1} \neq  P_{2}\}$; we can detect a difference in distribution with the sample size growing to infinity. Furthermore, the calibration strategy described above is also consistent in the same sense.   
\end{theorem}

\subsection{Missing at Random (MAR) mechanism}
\label{sec:mar}
We assume a MAR mechanism where the probability of being missing on the second time point is based on the corresponding
value on the first time point, which can be described as follows
\begin{equation}
\mathbf{X}=\underbrace{\begin{pmatrix}   X_{11} \\ X_{21} \end{pmatrix} \cdots    \begin{pmatrix}   X_{1n_1} \\ X_{2n_1} \end{pmatrix}}_{
\substack{\text{Complete data $\mathbf{X}^{\text{com}}$} \\ \text{$n_1$ observations}}}
\hspace{0.2cm}  \underbrace{\begin{pmatrix}   X_{1n_1+1} \\  -  \end{pmatrix} \cdots   \begin{pmatrix}   X_{1n_1+n_2} \\ -  \end{pmatrix}}_{
\substack{\text{Incomplete data $\mathbf{X}^{\text{inc}}_1$} \\ \text{$n_2$ observations}}}, \\
\end{equation}
where $n=n_1 + n_2$. We denote by $\pi\left(\cdot\right)= P\left(\delta_{2j}=1 |X_{1j}=\cdot\right)$, the conditional probability that the observation $X_{2j}$ will be missing given $X_{1j}$. A natural way to incorporate the missing data mechanism in the test statistic is to associate weight $\omega_j$ with the $j$-th observation via an inverse probability weighting (IPW) estimator \cite{tsiatis2007semiparametric}, given by 
\begin{equation} 
\omega_j= \frac{\delta_{2j}}{n\pi\left(X_{1j}\right)},~ j = 1,\dots, n.
\label{eq:weights}
\end{equation}
In practice, we estimate the probability $\pi(\cdot)$ by means of a binary classification algorithm. We denote by $\tilde{\omega}_j$ the estimated weight. We propose the following test statistic
\begin{align*}
\mathcal{T}\left(\mathbf{X}\right) &= \mathcal{T}\left(\mathbf{X}^{\text{com}}_1,\mathbf{X}^{\text{com}}_2\right) =\norm{\widetilde{\phi}^{\text{com}}_{1}-\widetilde{\phi}^{\text{com}}_{2}}_{\mathcal{H}}^{2}\\
&=  \langle \sum_{j=1}^{n_1} \tilde{\omega}_j  k\left(\cdot, X_{1j}\right)-\sum_{j=1}^{n_1} \tilde{\omega}_j k\left(\cdot, X_{2j}\right), \sum_{j=1}^{n_1} \tilde{\omega}_j  k\left(\cdot, X_{1j}\right)-\sum_{j=1}^{n_1} \tilde{\omega}_j k\left(\cdot, X_{2j}\right) \rangle \\ 
&= \sum_{i=1}^{n_1}\sum_{j=1}^{n_1} \tilde{\omega}_i \tilde{\omega}_j k\left(X_{1i},X_{2j}\right) + \sum_{i=1}^{n_1}\sum_{j=1}^{n_1} \tilde{\omega}_i \tilde{\omega}_j k\left(X_{1i},X_{2j}\right)- 2 \sum_{i=1}^{n_1}\sum_{j=1}^{n_1} \tilde{\omega}_i \tilde{\omega}_j k\left(X_{1i},X_{2j}\right).
\end{align*}

In this scenario, we propose to calibrate the test under the null hypothesis in an analogous manner to the MCAR mechanism. Specifically, for each bootstrap iteration we propose to use the following estimator
\begin{align*}
\mathcal{T}^{b}\left(\mathbf{X}\right) = \frac{1}{n_1^{2}}  \sum_{i=1}^{n_1}\sum_{j=1}^{n_1}  w^{b}_{i} w^{b}_{j} \tilde{\omega}_i \tilde{\omega}_j \left[ k\left(X_{1i},X_{1j}\right)+ k\left(X_{2i},X_{2j}\right)-2k\left(X_{1i},X_{2j}\right) \right].
\end{align*}
    
\subsection{Kernel choice and kernel hyperparameters}
\label{sec:kernel_choice}

We propose using the Gaussian kernel $k\left(X,Y\right)= e^{-\norm{X-Y}^{2}/\sigma^2}$ for $X,Y \in \mathbb{R}$, and $k\left(X,Y\right)= e^{-d^{2}_{\mathcal{W}_{2}}\left(X,Y\right)/ \sigma^2}$ for $X,Y \in \mathcal{D}$, where $\sigma>0$. Importantly, the Gaussian kernel is a characteristic kernel, and thus we can detect asymptotically any difference in distribution. The kernel bandwidth $\sigma$ was estimated through the median heuristic $\sigma ^2= median \{||X_i-X_j||^{2}:1\leq i< j\leq n \}$. 

\subsection{Simulation study}
\label{sec:simulation_study}

We investigate the finite sample behavior of the above methods in extensive simulations. A total of $2,000$ simulations were performed for both MCAR and MAR scenarios. Methods were examined with respect to their Type-I error rate control at level 5\%. A total of $2,000$ bootstrap runs and permutation replicas were held. The wild bootstrap parameter $l_{n_1}$ was selected according to $l_{n_1}= \sqrt{n_1}$.

The observations were generated by mimicking the sort of distributional representations commonly obtained from CGM data. Since the 2-Wasserstein distance depends only on quantile functions, observations were sampled from the following location-scale model on quantile functions \cite{petersen2021}: let $Z\in \mathbb{R}^{p}$ be a random vector of predictor variables and let $Q_0$ be a fixed quantile function; here we considered the age as the only predictor variable and fixed $Q_0\left(t\right)= 70+240t$ in the range of glucose values expected from type-2 diabetes; let $\eta\left(z\right)= a_0+a_1z_1$ and $\tau\left(z\right)= b_0+b_1z_1$ be the location and scale components of the model, respectively, where $a= \left(a_0,a_{1}\right)$ and $b= \left(b_0, b_{1}  \right)$ are the corresponding coefficients and we assume that $\tau\left(Z\right)>0$ almost surely; let $V_1$ and $V_2$ two random variables that satisfy $E\left(V_1|Z\right)=0, E\left(V_2|Z\right)=1$, and $V_2>0$ almost surely; the model is given by 
\begin{equation}
Q\left(t\right)= V_1+V_2\eta\left(Z\right)+V_2\tau\left(Z\right)Q_0\left(t\right),
\end{equation}

\subsubsection{MCAR scenario.}
\label{sec:sim_mcar}
We fixed $n_1=n_2=n_3=150$. In order to introduce correlation structure into the quantile functions for $\mathbf{X}^{\text{com}}_1$ and $\mathbf{X}^{\text{com}}_2$, we sampled variables $V_1^*$ and $V_2^*$ from bivariate uniform distributions with correlation given by $\rho\in \{0.00,0.20,0.40,0.60,0.80\}$. The location-scale model is given by $V_1=-20+40V_1^*$ and $V_2= 0.8+0.4V_2^*$, and fixed parameters $a_0=b_0=0$, $a_1=0.3$ and $b_1=0.005$. The observations for $\mathbf{X}^{\text{inc}}_1$ and $\mathbf{X}^{\text{inc}}_2$ were i.i.d. generated and then we applied the same location-scale model than before. A total of $2,000$ simulations were performed assuming that the age was distributed as $Z_1,Z_2\sim \mathcal{U}_{[30,50]}$ both at the beginning and at the end of the study, that is, for all the variables in $\mathbf{X}$. Another $2,000$ simulations were performed assuming that the age was distributed as $Z_1\sim \mathcal{U}_{[30,50]}$ at the beginning of the study, that is, for all the variables in $\mathbf{X}^{\text{com}}_1$ and $\mathbf{X}^{\text{inc}}_1$, and was distributed as $Z_2\sim \mathcal{U}_{[50,70]}$ at the end of the study, that is, for all the variables in $\mathbf{X}^{\text{com}}_2$ and $\mathbf{X}^{\text{inc}}_2$.

\subsubsection{MAR scenario.}
\label{sec:sim_mar}
We fixed $n=300$. The missing mechanism is given by $P\left(\delta_{2j}=1|Y_1,Y_2\right)= (1+e^{-1+ Y_1+ Y_2})^{-1}$, $j=1,\dots, n$, where $Y_1,Y_2\sim \mathcal{N}\left(0,1\right)$  are two independent random variables. We introduced correlation structure into the quantile functions for $\mathbf{X}^{\text{com}}_1$ and $\mathbf{X}^{\text{com}}_2$ as we did in the MCAR scenario. We used the same location-scale model. The same methodology as in the MCAR scenario was applied for sampling the age.

\subsubsection{Results.} 
Table \ref{table:tabla1} shows the results of the simulation study. We can see the test calibration under the null hypothesis is acceptable. However, there are some biases in the two situations due to missing data mechanisms. As the discrepancy of the null hypothesis increases, the test rejects the more null hypothesis, and in some cases, it is clear the consistency of the new methods with $100$ percent of reject cases.   

 \begin{table}[ht]
	\centering
	 \begin{tabular}{ccccc}
		  \hline
		  $\rho$ & $Z_1$ & $Z_2$ & MCAR & MAR    \\ 
	    \hline
		  $0.00$ & $\mathcal{U}_{[30,50]}$ & $\mathcal{U}_{[30,50]}$ & $0.03$ & $0.03$ \\ 
		  $0.20$ & $\mathcal{U}_{[30,50]}$ & $\mathcal{U}_{[30,50]}$ & $0.04$ & $0.03$ \\ 
		  $0.40$ & $\mathcal{U}_{[30,50]}$ & $\mathcal{U}_{[30,50]}$ & $0.05$ & $0.03$ \\ 
		  $0.60$ & $\mathcal{U}_{[30,50]}$ & $\mathcal{U}_{[30,50]}$ & $0.03$  & $0.04$  \\ 
		  $0.80$ & $\mathcal{U}_{[30,50]}$ & $\mathcal{U}_{[30,50]}$ & $0.04$ & $0.04$ \\ 
        \hline
          $0.00$ & $\mathcal{U}_{[30,50]}$ & $\mathcal{U}_{[50,70]}$ & $0.98$ & $0.88$ \\ 
		  $0.20$ & $\mathcal{U}_{[30,50]}$ & $\mathcal{U}_{[50,70]}$ & $0.99$ &  $0.90$ \\ 
		  $0.40$ & $\mathcal{U}_{[30,50]}$ & $\mathcal{U}_{[50,70]}$ & $0.99$ & $0.91$ \\ 
          $0.60$ & $\mathcal{U}_{[30,50]}$ & $\mathcal{U}_{[50,70]}$ & $0.99$ & $0.93$ \\ 
		  $0.80$ & $\mathcal{U}_{[30,50]}$ & $\mathcal{U}_{[50,70]}$ & $0.99$ & $0.96$  \\
		  \hline
		 \end{tabular}
	 \caption{The proportion of simulations rejecting the null hypothesis is shown under MCAR and MAR mechanisms.}
	 \label{table:tabla1}
\end{table}

\subsection{Paired missing data clustering}
\label{sec:cluster}

Let $\mathbf{X}=\left\{\left(X_{1j}, X_{2j}, \delta_{2j}\right)\right\}^{n}_{j=1}$, be a dataset of i.i.d. random variables obtained under a MAR mechanism, where we denote again by $\pi\left(\cdot\right)= P\left(\delta_{2j}=1 |X_{1j}=\cdot\right)$, the probability that the observation $X_{2j}$ will be missing. We associate a weight $\tilde{\omega}_j$ with the $j$-th observation via an IPW estimator, by applying equation (\ref{eq:weights}). Let $\mathbf{X}_j=(X_{1j},X_{2j})^{\top}$, $\mathbf{X}_h=(X_{1h},X_{2h})^{\top} \in \mathbf{X}^{\text{com}}$ be two different complete paired samples. We define the following bivariate kernel 
$k\left(\mathbf{X}_j,\mathbf{X}_h\right)= e^{-\left(d^{2}_{\mathcal{W}_2}(X_{1j},X_{1h})+d^{2}_{\mathcal{W}_2}(X_{2j},X_{2h})\right)/\sigma^2}$, where $\sigma^2= median\{d^{2}_{\mathcal{W}_2}\left(X_{1j},X_{1h}\right)+d^{2}_{\mathcal{W}_2}\left(X_{2j},X_{2h}  \right): 1\leq j<h\leq n \}$. 

Consider a disjoint partition $\mathbf{X}^{\text{com}}=\bigcup _{i=1}^k C_i$, with $C_i\cap C_l=\emptyset$, for all $i\neq l$. Following \cite{francca2021kernel}, we aim to build a new partition $\tilde{C}_1,\dots, \tilde{C}_k$ by maximizing an objective function given by
\begin{equation}
\left(\tilde{C}_1,\dots, \tilde{C}_k\right)= \arg \max_{\left(C_1,\dots,C_{k}\right)} \sum_{i=1}^{k} \frac{1}{v_i} \sum_{\mathbf{X}_{j},\mathbf{X}_{h} \in C_i} \tilde{\omega}_j \tilde{\omega}_h k\left(\mathbf{X}_{j},\mathbf{X}_{h}\right), 
\end{equation}
where $v_i= \sum_{\mathbf{X}_{j}\in \mathcal{C}_i} \tilde{\omega}_j$. We can iteratively solve this optimization problem by measuring the impact of moving each observation to another cluster. Let denote by $S_i= \sum_{\mathbf{X}_{j},\mathbf{X}_{h}\in C_i} \tilde{\omega}_j \tilde{\omega}_h k(\mathbf{X}_{j},\mathbf{X}_{h})$ the internal similarity of cluster $C_i$, and $S_i\left(\mathbf{X}_{j}\right)=  \sum_{\mathbf{X}_{h} \in C_i} \tilde{\omega}_j \tilde{\omega}_h k(\mathbf{X}_{j},\mathbf{X}_{h})$ the internal similarity with respect to the observation $\mathbf{X}_{j}$. By moving the observation $\mathbf{X}_{j}$ from cluster $C_i$ to $C_l$ we change the result of the objective function by
\begin{equation}
\Delta S^{i\to l} \left(X_{2j}\right)= \frac{S^{+}_{l}  }{v_l+ \tilde{\omega}_j} + \frac{S^{-}_{i}}{v_i- \tilde{\omega}_j}-\frac{S_{l}}{v_l}-\frac{S_{i}}{v_i},
\end{equation}
where $S_l^{+}= S_l+2S_l\left(\mathbf{X}_{j}\right)+\tilde{\omega}_j \tilde{\omega}_j k\left(\mathbf{X}_{j},\mathbf{X}_{j}\right)$ is the internal similarity of the new cluster $C_l$ after the addition of the observation $\mathbf{X}_{j}$, and $S_i^{-}= S_i-2S_i\left(\mathbf{X}_{j}\right)+\tilde{\omega}_j \tilde{\omega}_j k\left(\mathbf{X}_{j},\mathbf{X}_{j}\right)$ is the internal similarity of the new cluster $C_i$ after removing the observation $\mathbf{X}_{j}$. Ultimately, we compute $i^{*}= \arg \max_{l=1\dots,k|l\neq i} \Delta Q^{i\to l} \left(\mathbf{X}_{j}\right)$, and if $\Delta S^{i\to i^{*}}(\mathbf{X}_{j})>0$ we move $\mathbf{X}_{j}$ to cluster $C_{i^*}$, otherwise we keep it in $C_i$.

\section{Illustrative data analysis}
\label{sec:aegis}
As a practical application, we consider an ongoing longitudinal, population-based study by \cite{gude2017}, aimed at analyzing the evolution of different clinical biomarkers related to circulating glucose in a initial random sample of $1516$ patients over $10$ years. In addition, a CGM are performed every five years on a randomized subset of patients. Specifically, at the beginning of the study, $581$ participants were randomly selected for wearing a CGM device for $3$-$7$ days. Out of the total of $581$ participants, $68$ were diagnosed with diabetes before the study and $22$ during the first five years. Table \ref{tab:aegis_table} shows the baseline characteristics of these $581$ patients grouped by sex. After a five-year follow-up, only 161 participants agreed to perform a second glucose monitoring. 

\begin{table}[ht!]
	\centering
	\begin{tabular}{lll}
		\hline
		& Men $(n=220)$ & Women $(n=361)$ \\ 
		\hline
		Age, years & $47.8\pm 14.8$ & $48.2\pm14.5$ \\ 
		A1c, \% & $5.6\pm0.9$ & $5.5\pm0.7$ \\ 
		FPG, mg/dl & $97\pm23$ & $91\pm21$ \\ 
		HOMA-IR, mg/dl.$\mu$ IU/ml & $3.97\pm5.56$ & $2.74\pm2.47$ \\ 
		BMI, kg/m$^2$ & $28.9\pm4.7$ & $27.7\pm5.3$ \\ 
		CONGA, mg/dl  & $0.88\pm0.40$ & $0.86\pm0.36$ \\ 
		MAGE, mg/dl & $33.6\pm22.3$ & $31.2 \pm14.6$ \\ 
		MODD & $0.84\pm0.58$ & $0.77\pm0.33$ \\ 
		\hline
	\end{tabular}
	\caption{Baseline characteristics of AEGIS study participants with CGM monitoring by sex. Mean and standard deviation are shown.	A1c: glycated hemoglobin; FPG: fasting plasma glucose; HOMA-IR: homeostasis model assessment-insulin resistance; BMI: body mass index; CONGA: glycemic variability in terms of continuous overall net glycemic action; MAGE: mean amplitude of glycemic excursions; MODD: mean of daily differences.}
	\label{tab:aegis_table}
\end{table}

The AEGIS study raises some interesting questions that can be addressed with the present approach.

{\bf Changes in CGM profile with ageing}. Some recent works explore the important role of ageing in glucose dysregulation, and the difficulties inherent in maintaining glucose homeostasis as close to normal as possible \cite{chia2017}. The proposed $\mathcal{T}$-test gives us the opportunity to examine if there exist statistical differences after five years at a distributional level. We estimate the missing data mechanism by means of logistic regression, using as predictors the age and glycaemic status (normoglycemic, prediabetes or type-2 diabetes) at the beginning of the study and sex of each participant. We applied the $\mathcal{T}$-test considering glucodensities at both time points to check the null hypothesis of equality of distributions. We obtained a p-value = $0.048$, identifying significant differences at both time points.

\begin{figure}[ht!]
	\centering
	\includegraphics[width=0.7\linewidth]{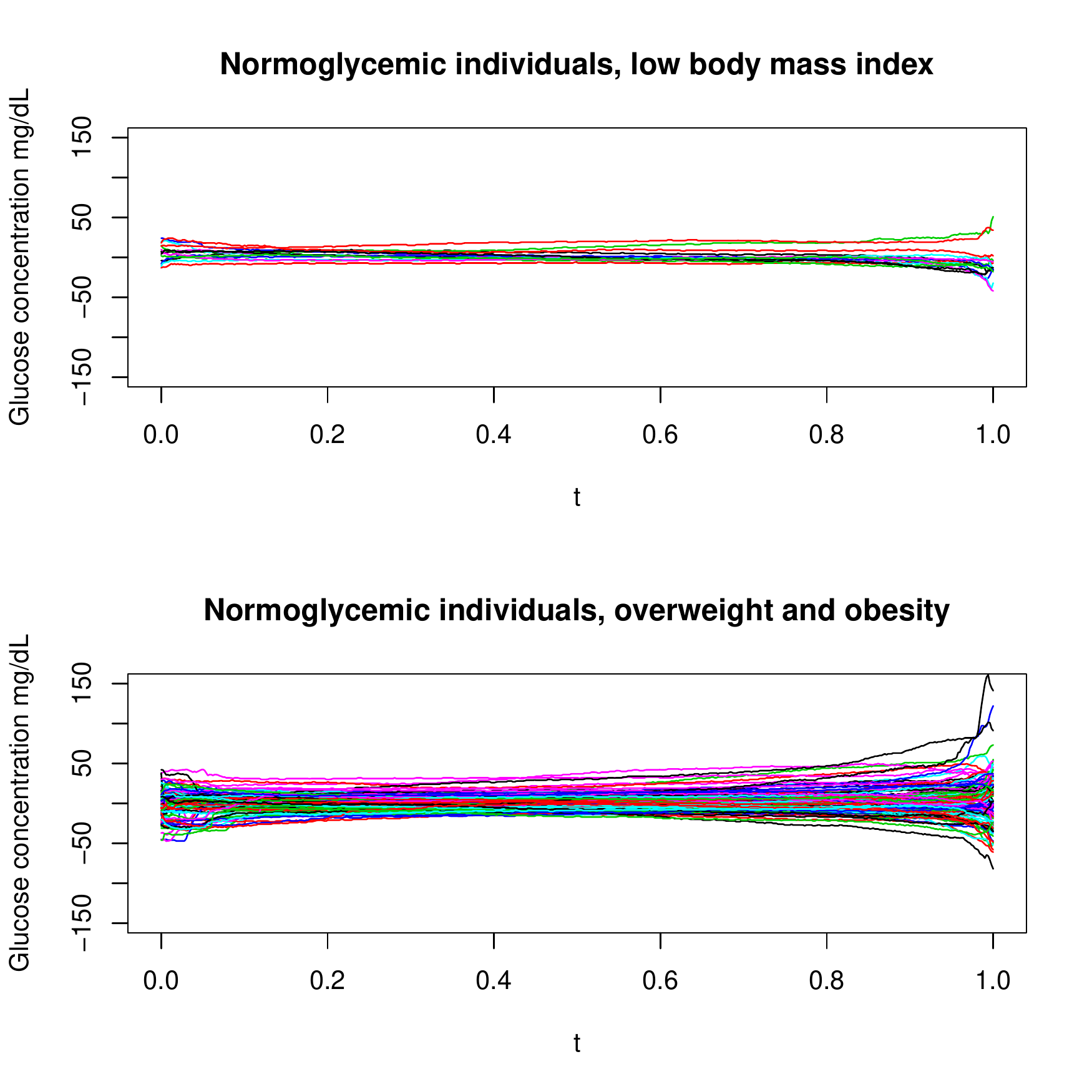}
	\caption{Difference between the quantile curves (before and after) in normoglycemic individuals according to body mass status. The dispersion is more significant for the overweight and obesity subgroup, consistent with an increasing glycemic risk.}
	\label{fig:obesity}
\end{figure}

{\bf Obesity in diabetes}. Obesity is a critical risk factor for the development of type-2 diabetes \cite{leong1999obesity}. In order to further characterize this risk subpopulation, we analyzed those normoglycemic subjects with overweight in the AEGIS dataset, by examining again if there exist statistical differences after five years at a distributional level. We applied the $\mathcal{T}$-test to check the null hypothesis in the following two subgroups of the normoglycemic population: i) individuals with a body mass index less than $22 Kg/m^{2}$ (low body mass index); ii) individuals with a body mass index higher than $22 Kg/m^{2}$ (overweight and obesity). In the first case we obtained a $p$-value = $0.36$, providing no evidence against the null hypothesis, while in the second case we obtained a $p$-value = $0.056$, which can be interpreted as borderline. Figure \ref{fig:obesity} shows the difference between the quantile curves in these two subgroups. 

\begin{figure}[ht!]
	\centering
	\includegraphics[width=0.7\linewidth]{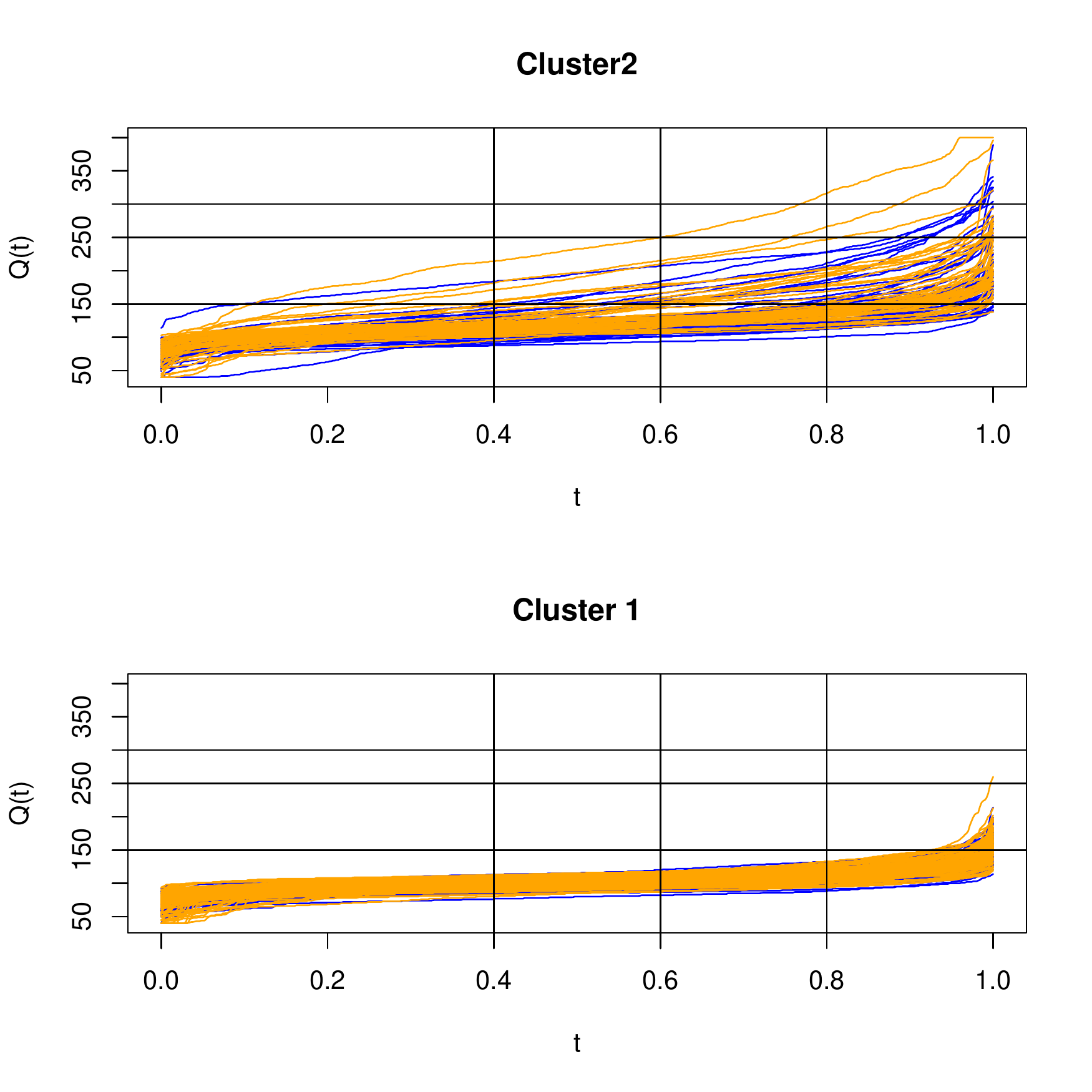}
	\caption{Resulting clusters are shown. Both quantile curves at the beginning of the study (blue) and five years later (orange) are shown for each cluster.}
	\label{fig:clustering}
\end{figure}

{\bf Patient stratification}. Clustering analysis can be a useful tool for providing distinctive and meaningful patient phenotypes and, consequently, in guiding patient stratification for delivering more personalized care \cite{kosorok2019precision}. We applied a clustering analysis to those individuals for whom CGM has been performed at both time points. Figure \ref{fig:clustering} shows the resulting two clusters. The individuals in cluster $1$ do not present significant changes between both time points, while some significant differences are noted in cluster $2$. Table \ref{table:groups} shows the baseline clinical characteristics of each cluster. Both groups of individuals have important differences in insulin resistance and glycaemic variability metrics. Importantly, in cluster 2 the average glycaemic characteristics in terms of glycated hemoglobin and fasting plasma glucose are consistent with prediabetes (5.7\% $\leq$ A1c $\leq$ 6.4\% or 100 mg/dl $\leq$ FPG $\leq$ 125 mg/dl according to American Diabetes Association guidelines). In contrast, cluster 1 is composed of normoglycemic individuals. Ultimately, clustering results effectively correlates with a significant change in the glycaemic status. 

Finally, we performed stepwise logistic regression with forward selection to identify which baseline characteristics independently predicted the corresponding group, resulting age, FPG and CONGA. We checked the null hypothesis that each coefficient is equal to zero. Table \ref{table:stepwise} shows the results of this analysis, identifying FPG and CONGA as the subset of characteristics that best predicted the outcome.

\begin{table}[ht!]
	\centering
	\begin{tabular}{lll}
		\hline
		& cluster 1 & cluster 2 \\ 
		\hline
		Age (years decimal) & $43.66\pm 12.80$ & $53.11\pm 12.34$ \\ 
		A1c, \% & $5.27\pm 0.25$ & $6.20 \pm 1.07$ \\ 
		FPG, mg/dl & $84.83 \pm 9.93$ & $108.39 \pm 32.61$ \\ 
		HOMA-IR, mg/dl.$\mu$ IU/ml& $2.28 \pm 1.16$ & $4.97 \pm 8.50$ \\ 
		BMI, kg/m$^2$ & $27.09 \pm 4.87$ & $29.29\pm 4.83$ \\ 
        Waist, cm & $87.29 \pm 13.60$ & $95.18 \pm 14.43$ \\ 
		CONGA, mg/dl & $0.75 \pm 0.20$ & $1.21 \pm 0.52$ \\ 
		MAGE, mg/dl & $26.16 \pm 7.16$ & $45.80\pm 24.58$ \\ 
		MODD & $0.66 \pm 0.18$ & $1.05 \pm  0.48$ \\ 
		\hline
	\end{tabular}

\caption{Clinical baseline characteristics for the individuals belonging to each cluster. Mean and standard deviation are shown.}
	\label{table:groups}
\end{table}

\begin{table}[ht!]
	\begin{center}
		\begin{tabular}{l c}
			\hline
			& Coefficients \\
			\hline
			(Intercept)                   & $-10.01$ $(1.82)^{***}$ \\
			age    & $0.04$  $(0.02)$       \\
			FPG    & $0.05$   $(0.02)^{**}$  \\
			CONGA & $3.60$  $(0.91)^{***}$ \\
			\hline
			& Quality measures \\
			\hline
			AIC                           & $140.37$       \\
			BIC                           & $152.69$       \\
			Log Likelihood                & $-66.18$       \\
			Deviance                      & $132.37$       \\
			\hline
			\multicolumn{2}{l}{ $^{***}p<0.001$; $^{**}p<0.01$}
			\end{tabular}
			\caption{Coefficients obtained from logistic regression. Results from some different model selection criteria for the fitted model are shown.}
					\label{table:stepwise}
			\end{center}

\end{table}

\section{Discussion}
\label{sec:discussion}

The analysis of paired data with missing values is becoming critical in longitudinal studies, particularly when comparing the participants' condition across different time points. The available methods in the literature are not applicable when data adopt non-vectorial representations, better suited to capture functional, structural or other complex forms of information increasingly common in current medicine. To overcome this limitation we have provided novel methods for hypothesis testing in the presence of complex paired missing data under both MCAR and MAR mechanisms. They are not based on any parametric assumption and use all observations within the matched pairs design. The methods are based on the notion of maximum mean discrepancy, a metric between mean embeddings in a RKHS that can be applied to both Euclidean and non-Euclidean data, with different structured, functional and distributional representations, by an appropriate design of the reproducing kernel. Specifically, the space of probability density functions has been used throughout the text to test the feasibility of this approach.

The asymptotic validity of the methods was proven and can be found in the appendix. In an extensive simulation study, the type-I error rate control of the tests has been examined under both MCAR and MAR mechanisms, performing well with different correlation coefficients. The sample size affects the behavior of the tests, since inference in a functional space customarily demands more data than in a vectorial space. Hence a worsening of performance is expected for very small sample sizes. 

The application of these methods to a real longitudinal, population-based, diabetes study has highlighted some of their capabilities and advantages to explore new clinical findings, by exploiting monitoring information along the continuous range of glucose values. It should be emphasized the robustness of the results, even in an scenario with an important proportion of missing data. Furthermore, a complementary clustering analysis has revealed the effectiveness of this approach to provide an early risk identification with the potential to enable a personalized strategy.

In order to simplify the application of these methods, they are freely available on GitHub
(https://github.com/).

\section*{Acknowledgments}

Funding for the project was provided by the .

\appendix

\end{document}